\documentstyle[12pt,aaspp4,amssym]{article}
\begin{document}
\def\Msun{M_\odot}
\def\et{{\rm et al.\ }}
\def\eg{{\rm e.g.\ }}
\def\ie{{\rm i.e.\ }}
\def\gs{\mathrel{\raise0.35ex\hbox{$\scriptstyle >$}\kern-0.6em 
\lower0.40ex\hbox{{$\scriptstyle \sim$}}}}
\def\ls{\mathrel{\raise0.35ex\hbox{$\scriptstyle <$}\kern-0.6em 
\lower0.40ex\hbox{{$\scriptstyle \sim$}}}}
\def\arcsper{\ifmmode \rlap.{''}\else $\rlap{.}''$\fi}
\def\arcmper{\ifmmode \rlap.{'}\else $\rlap{.}'$\fi}

\received{---}
\revised{---}
\accepted{---}
\journalid{---}{---}
\articleid{---}{---}
\slugcomment{Submitted to The Astrophysical Journal}


\title{Gravitational Lensing by NFW Halos}
\author{Candace Oaxaca Wright \& Tereasa G. Brainerd}
\affil{Boston University, Department of Astronomy, Boston, MA 02215}

\begin{abstract}

We investigate the gravitational lensing properties of dark matter halos
with Navarro, Frenk \& White (NFW) density profiles and derive an
analytic expression
for the radial dependence of the shear, $\gamma(x)$, due to these objects.
In addition, we derive an expression for the mean shear interior
to a given radius, $\overline{\gamma}(x)$, and use this
to quantify systematic errors that will arise in weak lensing 
mass estimates of astronomical objects in the case that
the mass estimate is based on an
{\it a priori} assumption that the underlying potential is that of
a singular isothermal sphere when, in fact, the potential is
that of an NFW--type object.  On mass scales between $10^{11} \Msun \ls M_{200}
\ls 10^{15} \Msun$, 
the assumption of an isothermal sphere potential
results in an overestimate of the halo mass, and the amount
by which the mass is overestimated increases linearly with the value
of the NFW concentration
parameter.  
Considerable overestimates of the mass ($\sim 60\%$) can occur
for galaxy--sized halos, but for rich clusters the mass overestimate is
small.
The degree to which the mass is systematically in error
is dependent upon the cosmology adopted, with a COBE-normalized standard
CDM model yielding the largest systematic errors for a given true
value of the halo mass.

\end{abstract}


\section{Introduction}

Several recent numerical investigations
(e.g., Navarro, Frenk \& White 1997, 1996, 1995)
have indicated the existence of a universal
density profile for dark matter halos that results from the generic
dissipationless collapse of density fluctuations.
Interior to the virial radius, the Navarro, Frenk \& White (NFW)
profile appears to be a very good description of the
radial mass distribution of 
simulated objects that span 9 orders of magnitude in mass  (mass scales
ranging from that of globular clusters to that of large galaxy clusters).
The apparent generality of the NFW density profile has been confirmed
independently by a number of studies
(e.g., Bartelmann et al. 1998; Thomas et al. 1998, 
Carlberg et al. 1997; Cole \& Lacey 1997; 
Kravtsov, Klypin \& Khokhlov 1997;
Tormen, Bouchet \& White 1997); however, there are a few controversial
claims that the NFW prescription may fail at very small radii 
(e.g., Ghigna et al. 1998; Moore et al. 1998).

The NFW density profile is given by 
\begin{equation}
\rho(r) = 
\frac{\delta_{c}\rho_{c}}{\left(r/r_{s}\right)
\left(1+r/r_{s}\right)^{2}},
\label{rhonfw}
\end{equation}
where $\rho_{c} = \frac{3H^{2}(z)}{8\pi G}$ is the critical density for 
closure of the universe
at the redshift, $z$, of the halo, $H(z)$ is Hubble's 
parameter at that same redshift, and  $G$ is Newton's constant.
The scale radius $r_{s} = r_{200}/c$ is a characteristic radius of 
the cluster, 
$c$ is a dimensionless number known as the concentration parameter, and 
\begin{equation}
\delta_{c}= \frac{200}{3}\frac{c^{3}}{\ln(1+c)-c/(1+c)}
\end{equation}
is a characteristic overdensity for the halo.  
The virial radius, $r_{200}$, is defined as 
the radius inside which the mass density of the halo is equal to
$200 \rho_c$
(see, e.g., Navarro, Frenk \& White 1997).
The mass of an NFW halo contained within a radius of $r_{200}$
is therefore
\begin{equation}
M_{200}  \equiv M(r_{200}) =  \frac{800\pi}{3} \rho_c r_{200}^{3} 
   =  \frac{800\pi}{3}
\frac{\overline{\rho}(z)}{\Omega(z)} r_{200}^3
\label{m200}
\end{equation}
where $\overline{\rho}(z)$ is the mean mass 
density of the universe at redshift $z$ and 
$\Omega(z)$ is the density parameter at redshift $z$.

Although it has not been proven categorically, it is certainly
widely--thought that
the masses of large galaxies, groups of galaxies, galaxy
clusters, and superclusters
are dominated by some form of dissipationless dark matter.  Therefore, it
would not be unreasonable to expect that the spherically--averaged density
profiles of these objects would be approximated fairly well by NFW profiles.
Observationally, the total masses and mass--to--light ratios of
these objects are not constrained especially well at present; however, this
situation is changing rapidly, due in large part to the fact that
high--quality imaging of gravitational lens systems is yielding 
direct constraints on the nature of the mass distribution within
the dark matter halos.  

Observations of gravitational lensing provide powerful
constraints on both the total mass and the mass distribution within
the lens itself, owing to the fact that one essentially uses
photons emitted by objects more distant than the lens
to trace the underlying gravitational potential of the lens directly.
In particular, large clusters of galaxies (which are both massive and 
centrally--condensed) are especially
good gravitational lens candidates,
and detections of the coherent pattern of
weak lensing shear due to a number of clusters
has led to interesting constraints on the masses of these objects
(e.g., Tyson, Wenk \& Valdes 1990;
Bonnet et al.\ 1994; Dahle, Maddox \& Lilje 1994; Fahlman et al.\ 1994;
Mellier et al.\ 1994; Smail et al.\ 1994, 1995, 1997;
Tyson \& Fischer 1995; Smail \& Dickinson 1995; Kneib et al\ 1996;
Seitz et al.\ 1996; Squires et al.\ 1996ab; Bower \& Smail 1997;
Fischer et al.\ 1997; Fischer \& Tyson 1997; Luppino \& Kaiser 1997;
Clowe et al.\ 1998; Hoekstra et al.\ 1998).  Although more
controversial than the results for lensing clusters, detections of
systematic weak lensing of distant field galaxies by foreground 
field galaxies have
been reported and these have been used to place constraints on
 the physical sizes
and total masses of the dark matter halos of the lens galaxies
(e.g., Brainerd, Blandford \& Smail 1996; Griffiths et al.\ 1996;
Ebbels 1998; Hudson et al.\ 1998; Natarajan et al.\ 1998).  Additionally,
a detection of the coherent weak lensing shear due to a supercluster has
been reported recently (Kaiser et al.\ 1998).

Because of the apparent direct applicability of the NFW density profile to
the dominant mass component of all of these objects, and because of
the potential of observations of gravitational lensing to provide strong,
direct constraints on the amount and distribution
of dark matter within them,
we investigate the lensing
characteristics of dark matter halos with generic NFW--type density profiles
in this paper. In \S2
we compute the convergence and the shear profiles 
of NFW halos.  In \S3 we compare the mean shear
induced by NFW lenses to that of simpler singular isothermal sphere 
(SIS) lenses and consider the implications of our results for 
possible systematic errors in lens masses that are determined 
in observational investigations which invoke an {\it a priori} assumption of
an isothermal lens potential. 
A discussion of the results is presented in \S4.

\section{Convergence and Shear of an NFW Object}

We perform all of our calculations below using the 
thin lens approximation, in which an object's lensing properties can be
computed solely from a scaled, 2-dimensional Newtonian potential. The thin
lens approximation is 
valid in the limit that the scale size of the lens is very 
much less than the path length traveled by the photons as they propagate from
the source to the lens and from the lens to the observer.
In this case the lensing 
properties of an object are completely described by two quantities, 
the convergence, $\kappa$, and the shear, $\vec{\gamma}$. 
The names of these quantities are indicative of their effects upon
a lensed image; the convergence describes the isotropic focusing
of light rays while the shear describes the effect of tidal gravitational
forces.  Convergence acting alone
leads to an isotropic magnification or demagnification while
the shear induces distortions in the shapes of lensed images.

If we define $z$ to be the optic axis, then for a lens with a 3-dimensional
potential $\Phi(D_d\vec{\theta},z)$ we can formulate a 
conveniently--scaled potential as projected on the sky:
\begin{equation} 
\psi(\vec{\theta}) = \frac{D_{ds}}{D_{d}D_{s}}\frac{2}{c^{2}}\int 
\Phi(D_{d}\vec{\theta}, z) dz.
\end{equation}
Here $\vec{\theta}$ is a radius vector on the sky and
$D_d$, $D_s$, and $D_{ds}$ are, respectively,
the angular diameter distances between
the observer and the lens, the observer and the source, and the
lens and the source.  Under the definition of 
$\psi(\vec{\theta})$ above, the 
convergence and the components of the shear tensor may be written
as straightforward combinations of second-order 
derivatives of $\psi$ with respect to image plane coordinates
$\vec{\theta} =(\theta_{1}, \theta_{2})$,
\begin{equation}
\kappa(\vec{\theta}) = \frac{1}{2}\left(
\frac{\partial^{2}\psi}{\partial\theta_{1}^{2}} + 
\frac{\partial^{2}\psi}{\partial\theta_{2}^{2}} \right)
\end{equation}
\begin{equation}
\gamma_{1}(\vec{\theta}) =  \frac{1}{2}\left(
\frac{\partial^{2}\psi}{\partial\theta_{1}^{2}} -
\frac{\partial^{2}\psi}{\partial\theta_{2}^{2}} \right) 
\end{equation}
\begin{equation}
\gamma_{2}(\vec{\theta}) = \frac{\partial^{2}\psi}
{\partial\theta_{1}\partial\theta_{2}} =\frac{\partial^{2}\psi}
{\partial\theta_{2}\partial\theta_{1}}.
\end{equation}
The magnitude of the shear is simply
$\gamma = |\vec{\gamma}| = \sqrt{\gamma_{1}^{2}
+ \gamma_{2}^{2}}$ (e.g., Schneider, Ehlers \& Falco 1992).
In the limit of
weak gravitational lensing, the convergence and
shear are formally small (i.e., $\kappa << 1$, $\gamma << 1$),
the ellipticity induced in the image of an intrinsically
circular source due to lensing
is of order $\gamma/2$, and the position angle
of the lensed image ellipse is of order the phase of
$\vec{\gamma}$ (e.g., Schramm \& Kayser 1995; Seitz \& Schneider 1997).

The local value of the
convergence may be expressed simply as the ratio of the local value of
the surface
mass density to the critical surface mass density:
\begin{equation}
\kappa(\vec{\theta}) = \frac{\Sigma(\vec{\theta})}{\Sigma_c},
\end{equation}
where
\begin{equation}
\Sigma_c \equiv \frac{c^2}{4\pi G} \frac{D_s}{D_d D_{ds}}
\end{equation}
(e.g., Schneider, Ehlers \& Falco 1992) and $c$ in the equation
above is the velocity of light.
The radial dependence of the surface mass density of a
spherically symmetric lens such as an NFW lens is obtained simply 
by integrating  
the 3-dimensional density profile along the line of sight, 
\begin{equation}
\Sigma(R) = 2 \int_{0}^{\infty}\rho(R,z)dz,
\end{equation}
where $R = D_d \sqrt{\theta_1^2 + \theta_2^2}$ is the projected radius
relative to the center of the lens.

For convenience we will adopt a dimensionless radial distance,
$x = R/r_{s}$.  Integrating equation (\ref{rhonfw}) along the line of sight,
the radial dependence of the
surface mass density of an NFW lens can then be written as:
\begin{equation}
\Sigma_{\rm nfw}(x) = \left\{ \begin{array}{ll}
\frac{2r_{s}\delta_{c}\rho_{c}}{\left(x^{2}-1\right)}
\left[1-\frac{2}{\sqrt{1-x^{2}}}{\rm arctanh}\sqrt{\frac{1-x}{1+x}}\hspace{0.15cm}
 \right] 
& \mbox{$\left(x < 1\right)$} \\ 
 & \\
\frac{2r_{s}\delta_{c}\rho_{c}}{3} & \mbox{$\left(x = 1\right)$} \\ 
 & \\
\frac{2r_{s}\delta_{c}\rho_{c}}{\left(x^{2}-1\right)}
\left[1-\frac{2}{\sqrt{x^{2}-1}}\arctan\sqrt{\frac{x-1}{1+x}}\hspace{0.15cm}
 \right] 
& \mbox{$\left(x > 1\right)$} 
\end{array}
\right.
\end{equation}
\noindent
(e.g., Bartelmann 1996). The radial dependence of the
convergence due to an NFW lens is then simply
$\kappa_{\rm nfw}(x) = \Sigma_{\rm nfw}(x)/\Sigma_c$.

Since the NFW density profile is spherically symmetric, the radial dependence
of the shear can be written as
\begin{equation}
\gamma_{\rm nfw}(x) = \frac{\overline{\Sigma}_{\rm nfw}(x) - \Sigma_{\rm nfw}(x)}{\Sigma_{c}}
\end{equation}
(e.g., Miralda--Escud\'e 1991) where $\overline{\Sigma}_{\rm nfw}(x)$ is
the mean surface mass density interior to the dimensionless
radius $x$.  In terms of this
radius, then, the mean surface mass density of an NFW 
halo is given by

\begin{equation}
\overline{\Sigma}_{\rm nfw}(x)= \frac{2}{x^2} \int_0^x x' 
\Sigma_{\rm nfw}(x') dx' = \left\{ \begin{array}{ll}
\frac{4}{x^{2}}r_{s}\delta_{c}\rho_{c}\left[
\frac{2}{\sqrt{1-x^{2}}}{\rm arctanh}\sqrt{\frac{1-x}{1+x}}+\ln\left(\frac{x}{2}
\right)\right]
& \mbox{$\left(x < 1\right)$} \\
 & \\
4r_{s}\delta_{c}\rho_{c}\left[1+\ln\left(\frac{1}{2}\right)\right]& 
\mbox{$(x = 1)$} \\
 & \\
\frac{4}{x^{2}}r_{s}\delta_{c}\rho_{c}\left[
\frac{2}{\sqrt{x^{2}-1}}\arctan\sqrt{\frac{x-1}{1+x}}+\ln\left(\frac{x}{2}
\right)\right] &
\mbox{$\left(x > 1\right)$}
\end{array}
\right. 
\end{equation}
and the radial dependence of the shear is, therefore, 
\begin{equation}
\gamma_{\rm nfw}(x) = \left\{ \begin{array}{ll}
\frac{r_{s}\delta_{c}\rho_{c}}{\Sigma_{c}}
g_{<}(x) & \mbox{$\left(x < 1\right)$} \\
 & \\
\frac{r_{s}\delta_{c}\rho_{c}}{\Sigma_{c}}
\left[\frac{10}{3} + 4\ln\left(\frac{1}{2}\right)\right] & \mbox{$\left(x = 1\right)$} \\
 & \\
\frac{r_{s}\delta_{c}\rho_{c}}{\Sigma_{c}}
g_{>}(x) & \mbox{$\left(x > 1\right)$}
\end{array}
\right. 
\label{gnfw}
\end{equation}
where the functions $g_{<,>}(x)$ above
depend upon only the dimensionless
radius $x$ and are explicitly independent of the cosmology:
\begin{eqnarray}
\hspace{-0.4cm}g_{<}(x) & \hspace{-0.15cm}= \hspace{-0.15cm}& \frac{8{\rm arctanh}
\sqrt{\frac{1-x}{1+x}}}{x^{2}\sqrt{1-x^{2}}}
+ \frac{4}{x^{2}}\ln\left(\frac{x}{2}\right) - \frac{2}{\left(x^{2}-1\right)}+
\frac{4{\rm arctanh}\sqrt{\frac{1-x}{1+x}}}{\left(x^{2}-1\right)\left(1-x^{2}\right)^{1/2}}  \label{gless} \\
\hspace{-0.4cm}g_{>}(x) &\hspace{-0.15cm} =\hspace{-0.15cm} & \frac{8\arctan\sqrt{\frac{x-1}{1+x}}}{x^{2}\sqrt{x^{2}-1}}
\hspace{0.1cm}
+ \hspace{0.1cm}\frac{4}{x^{2}}\ln\left(\frac{x}{2}\right) - \frac{2}{\left(x^{2}-1\right)}
\hspace{0.1cm}+\hspace{0.1cm}
\frac{4\arctan\sqrt{\frac{x-1}{1+x}}}{\left(x^{2}-1\right)^{3/2}}.
\label{ggtr}
\end{eqnarray}
Equation (14) above can also be obtained straightforwardly from equations
(7) through (11) of Bartelmann (1996).
The radial dependence of the shear due to an NFW lens is shown in 
Fig.~1.

The shear due to a given lens (e.g., a cluster of galaxies) is
computed directly from the coherent distortion pattern that it induces
in the images of distant source galaxies.  In the realistic observational
limit of weak shear
and a finite number of lensed images, a measurement of the
mean shear interior to a radius $x$ centered on the
center of mass of the lens (i.e., $\overline{\gamma}(x)$)
is more easily determined than
the differential radial dependence of the shear (i.e., $\gamma(x)$).  
In the case of the
NFW profile, the mean shear interior
to a (dimensionless) radius $x$ can be computed directly
from equation (\ref{gnfw}) above:
\begin{equation}
\overline{\gamma}_{\rm nfw}(x) = \frac{2}{x^2} \int_{0}^{x} x' \gamma(x') dx' =
\frac{r_{s}\delta_{c}\rho_{c}}{\Sigma_{c}}
\left[\frac{2}{x^{2}}\left(\int_{0}^{1}g_{<}(x')x'dx'+
\int_{1}^{x}g_{>}(x')x'dx'\right)\right] .
\end{equation}

A useful fiducial radius interior to which one might measure the 
mean shear is the virial radius, $R=r_{200}$, or equivalently, interior to
$x = \left(r_{200}/r_{s}\right) = c$, where $c$ is the concentration 
parameter. 
For all masses of astrophysical interest $c$ is greater than 1 and, therefore,
the mean shear interior to the virial radius becomes
\begin{equation}
\overline{\gamma}_{\rm nfw}(r_{200}) = 
\frac{r_{s}\delta_{c}\rho_{c}}{\Sigma_{c}}
\left[\frac{2}{c^{2}}\left(\int_{0}^{1}
g_{<}(x')x'dx'+\int_{1}^{c}g_{>}(x')x'dx'\right)\right]
\end{equation}
which we rewrite as 
\begin{equation}
\overline{\gamma}_{\rm nfw}(r_{200}) = 
\frac{r_{s}\rho_{c}}{\Sigma_{c}}{\cal F}\left(c\right),
\label{gbarnfw}
\end{equation} 
where
\begin{equation}
{\cal F}\left(c\right) = \delta_{c}
\left[\frac{2}{c^{2}}\left(\int_{0}^{1}
g_{<}(x')x'dx'+\int_{1}^{c}g_{>}(x')x'dx'\right)\right]
\end{equation}
is a function of the concentration parameter alone.

\section{Comparison to the Singular Isothermal Sphere}

Like the NFW mass profile, the singular isothermal sphere (SIS) mass 
profile is characterized by a single parameter (i.e., the velocity dispersion,
$\sigma_v$).
The mass of an SIS interior to 
a three dimensional radius $r$ is:
\begin{equation}
M(r) = \frac{2\sigma_{v}^{2}r}{G}
\label{msis}
\end{equation}
(e.g., Binney \& Tremaine 1987) and the
mean gravitational lensing
shear interior to a radius $R$ that is induced by an SIS lens
is:
\begin{equation}
\overline{\gamma}_{\rm sis}\left(R\right) = 
\frac{1}{\Sigma_{c}}\frac{\sigma_{v}^{2}}{G R}
\label{gbarsis}
\end{equation}
(e.g.\ Schneider, Ehlers \& Falco 1992).

Because of its simplicity, the SIS density profile is sometimes adopted
in observational investigations in order to obtain an estimate
of the mass of a lens without fully reconstructing its true underlying
density profile (e.g., Tyson, Wenk \& Valdes 1990; Bonnet et al.\ 1994; 
Smail et al.\ 1994,
1997; Smail \& Dickinson 1995;  Bower \& Smail 1997; Fischer \& Tyson 1997).
By assuming
that the underlying potential of the lens is well--approximated
by an SIS, a measurement of the mean shear interior to a projected
radius $R$ leads directly to a measurement of the velocity dispersion
of the lens (e.g., equation \ref{gbarsis}), which in turn leads directly to an
estimate of the mass of the lens (e.g., equation \ref{msis}).

The NFW density profile, which is shallower than isothermal on small scales,
and which turns over to isothermal on large scales 
has, however, been shown to be a far better 
approximation than the SIS to the spherically--averaged density profiles of  
halos formed via dissipationless collapse.  Therefore, it is likely
that lens mass estimates based on an {\it a priori} assumption of an isothermal
potential will be systematically in error.  In this section we compare
the mean shear induced by NFW lenses to that induced by SIS lenses,
under the constraint that the NFW and SIS lenses both have identical 
virial radii, $r_{200}$, and, therefore, identical masses
interior to $r_{200}$.  From this we will then investigate the
possible systematic 
errors in lens mass estimates that would arise due to the assumption of
an isothermal potential when, in fact, the lens is best represented
by an NFW density profile. 

Let us consider two lenses which have identical masses, $M_{200}$,
interior to the
virial radius.  One of the lenses has an NFW
density profile with a concentration parameter of $c$ and the other
is a singular isothermal sphere with velocity dispersion
$\sigma_v$.  If these two objects have identical redshifts, $z_d$, and
act as lenses for populations of
source galaxies which have identical redshifts, $z_s$, then
from equations (\ref{gbarnfw}) and (\ref{gbarsis}) above, the
ratio of the mean shears induced by these two lenses interior to $r_{200}$ is
given by:
\begin{equation}
\frac{\overline{\gamma}_{\rm nfw}\left(r_{200}\right)}
{\overline{\gamma}_{\rm sis}\left(r_{200}\right)} 
= 
\frac{r_{s}\rho_{c}r_{200}}{\sigma_{v}^{2}}
G {\cal F}\left(c\right).
\label{gratio1}
\end{equation}
Using equations (\ref{m200}) and (\ref{msis}) above and recalling
that the concentration parameter is $c=r_{200}/r_s$, it is straightforward
to show that equation (\ref{gratio1}) reduces to
\begin{equation}
\frac{\overline{\gamma}_{\rm nfw}\left(r_{200}\right)}
{\overline{\gamma}_{\rm sis}\left(r_{200}\right)} =
\frac{3}{400\pi}\frac{{\cal F}\left(c\right)}{c},
\label{gratio2}
\end{equation}
which is a function solely of the concentration parameter of the
NFW lens and is explicitly independent of the redshift of the
sources, $z_s$.  Because of the dependence of the concentration parameter
on both the redshift of the lens
and the cosmology through $\overline{\rho}(z_d)$,
equation (\ref{gratio2}) is not explicitly independent of either
the cosmology or the lens redshift, $z_d$. However, for lenses of a
given mass, its dependence on
both $z_d$ and the cosmology is relatively weak.

Shown in Figs.\ 2 and 3 are the ratio of the mean shears interior to
$r_{200}$ for NFW and SIS lenses with virial masses in the range
of $10^{11} M_\odot \le M_{200} \le 10^{16} M_\odot$.  Fig. 2 shows
the results for lenses located at $z_d = 0.1$ and Fig. 3 shows the
results for lenses located at $z_d = 0.5$.  The four panels in 
the figures show the effects of varying the cosmology, and plotted along
the top axes of all of the panels is the NFW concentration parameter
which corresponds to the lens mass plotted on the lower axes.

Two of the cosmologies illustrated in Figs. 2 and 3 are standard
cold dark matter (CDM) cosmologies, which differ from one another only in the
choice of the normalization of the power spectrum (SCDM--I is a cluster
abundance normalization while SCDM--II is COBE-normalized).  The other
two cosmologies are an open CDM model with zero cosmological
constant (OCDM) and a spatially flat, low matter density CDM 
model with a large cosmological constant ($\Lambda$CDM).  The 
parameters adopted for each of the models are summarized in Table 1
where $\Lambda_0 = \lambda/3H_{0}^{2}$, $H_0 = 100h$~km/s/Mpc, $n$ is 
the index of the primordial power spectrum of density fluctuations and
\begin{equation}
\sigma_8 \equiv
\left< \left[ \frac{\delta\rho}{\rho}(8h^{-1} {\rm Mpc})
\right]^2 \right>^\frac{1}{2} .
\end{equation}

\vspace{0.5in}
\centerline{Table 1: Cosmological Model Parameters}
\vspace{0.1in}
\begin{center}
\begin{tabular}{|c|l|l|c|c|c|} \hline \hline
          & $\Omega_{0}$ & $\Lambda_{0}$ & $h$ & $\sigma_{8}$ & $n$ \\ \hline
SCDM--I   & 1.0 & 0.0 & 0.50 & 0.63 & 1.0 \\ \hline
SCDM--II  & 1.0 & 0.0 & 0.50 & 1.20 & 1.0 \\ \hline
OCDM      & 0.25 & 0.0 & 0.70 & 0.85 & 1.0 \\ \hline
$\Lambda$CDM& 0.25 & 0.75 & 0.75 & 1.30 & 1.0 \\ \hline
\hline
\end{tabular}
\end{center}
\vspace{0.2in}

The FORTRAN program {\em charden.f}, written 
and generously provided by Julio Navarro, was used to calculate
the values of the concentration parameters for the NFW lenses in the
above cosmologies. 
For each of the
cosmologies, $c$ was determined for
halos with masses in the range of $10^{11} M_{\odot} \le M_{200}
\le 10^{16} M_\odot$ at redshifts of $z_d = 0.1$ and $z_d = 0.5$.
These values of $c$ were then used in conjunction with equation
(\ref{gratio2}) to compute the ratio of the NFW to SIS mean shear
interior to the virial radius.  

For a given cosmology, it is clear
by comparing Fig.\ 2 with Fig.\ 3 that equation (\ref{gratio2}) is
only weakly dependent on the lens redshift, $z_d$.
The largest difference between the various panels in Figs.\ 2
and 3 which correspond to identical cosmologies occurs 
for SCDM lenses with masses $\sim 10^{11} \Msun$, and in this case
the difference between $z_d =0.1$ and $z_d = 0.5$
is only $\sim 10\%$.  Similarly, by comparing the 
results plotted in all of the individual panels of Fig.\ 2 and
Fig.\ 3 at fixed $z_d$, it is clear that equation (\ref{gratio2}) is
not tremendously sensitive to the cosmology.  In particular, the
$\Lambda$CDM, OCM, and
SCDM--I models all yield functions with nearly identical amplitudes
for a given value of $z_d$.
The SCDM--II model yields a function which is somewhat higher than
the other three models, exceeding the others by $\sim 25\%$ for
halos with masses $\sim 10^{11} \Msun$ and by $\sim 20\%$ for halos
with masses $\sim 10^{16} \Msun$.

Over the majority of the mass range investigated, the 
NFW lenses give rise to a mean shear interior to $r_{200}$ which
is systematically larger than that of the SIS lenses.  As a result,
if one were to measure the mean shear interior to a radius of
$r_{200}$ of an NFW halo, yet assume it to be an isothermal
sphere, the resulting estimate of the virial mass of the lens 
($M_{200}$) would be systematically
high.  From equations (\ref{msis}) and (\ref{gbarsis}) above, it
follows that the mass of an SIS lens interior to $r_{200}$ is 
simply:
\begin{equation}
M_{200}  = 2\Sigma_c r_{200}^2 \overline{\gamma}(r_{200})
\label{msis2}
\end{equation}
so that the mass inferred for the lens scales linearly with the
mean shear.  Therefore, the systematic error in the true virial mass
of the lens is simply the ratio of the mean shear due to an NFW lens
to that of an SIS lens with an identical amount of mass contained
inside $r_{200}$ (i.e., Figs.\ 2 and 3).

Shown in Fig.\ 4 is the ratio of the mean shear (interior to
$r_{200}$) of an NFW lens and an SIS lens, plotted as a function
of the NFW concentration parameter.  (As in Figs.\ 2 and 3, 
both lenses have identical masses interior to $r_{200}$).  From this 
figure, then, if one
were to measure a mean shear for a given NFW lens, yet model the 
lens as an isothermal sphere, the degree of systematic error in an
estimate of the virial mass would clearly be a 
function of the concentration parameter of the lens.  For a given
halo mass, the concentration parameter is a function of the cosmology
(e.g., the top axes of Figs.\ 2 and 3); 
however, it is always the case that for a
given cosmology, the larger the value of $c$, the lower is the
value of $M_{200}$.   The general conclusions that can be drawn
from Fig.\ 4 are: [1] the lower the mass of an NFW halo, the larger
the systematic error in the mass estimate if the lens is assumed
to be an isothermal sphere
and [2] for a halo of a given mass, the largest systematic
error in the mass estimate occurs in a COBE-normalized cosmology
(i.e., SCDM-II).
With the exception of SCDM-II for which the error is somewhat larger,
the systematic error in an
estimate of $M_{200}$ for rich clusters ($M_{200} \sim 10^{15} \Msun$)
is negligible ($\ls 10\%$).  The systematic error in an estimate
of $M_{200}$ for galaxy--mass objects ($M_{200} \sim 10^{11} \Msun$) is,
on the other hand, considerable (of order 55\% to 65\% for the
SCDM--II model and of order 30\% to 40\% for the other models).

\section{Discussion}

It is generally thought that the masses of large galaxies, galaxy
groups, galaxy clusters, and superclusters are dominated by some
form of dissipationless dark matter and, thus, it is not unreasonable to expect
that their underlying mass density profiles will be represented reasonably
well by NFW profiles.  In addition, since their total masses and
mass-to-light ratios are not strongly constrained at present, 
a significant effort is currently being devoted to the use of observations
of gravitational lensing by these objects to quantify
the amount and distribution
of dark matter within them.
We have, therefore, investigated the properties of NFW lenses in this paper and
we have presented analytic expressions for the radial
dependence of the convergence, $\kappa(x)$, and shear,
$\gamma(x)$, due to dark matter halos
which have NFW density profiles.  We have also presented an expression for the
mean shear interior to a given radius, 
$\overline{\gamma}(x)$, due to NFW lenses
and we have compared the mean shear interior to the virial
radius of an NFW lens to that yielded by a singular isothermal sphere
lens with an identical virial mass.

It is not uncommon for the mass of a gravitational lens to be estimated
under an assumption that the lens may be approximated by a singular
isothermal sphere.  However, it has been clearly demonstrated that the NFW
density profile is a far better approximation to the density profile
of objects formed by generic dissipationless collapse than is the
isothermal sphere.  We have computed the systematic error that would
be encountered in an estimate of the mass of an NFW lens, where the
lens is assumed {\it a priori} to be an isothermal sphere.  Over
mass scales of $10^{11} \Msun \ls M_{200} \ls 10^{15} \Msun$, the mass of the
NFW lens is systematically overestimated when it is assumed that, for
a given measured value of $\overline{\gamma}(r_{200})$, the lens can
be approximated by an isothermal sphere.

The size of the systematic error in the lens mass due to the isothermal sphere
assumption is a function of the NFW concentration parameter of the
lens, with the largest error occurring for halos with the largest
values of $c$ and, hence, with the smallest masses.  The systematic
error in the mass is not dramatic (i.e., not even as much as a
factor of $\sim 2$), but this is unsurprizing since the shape of the
NFW density profile in the outer regions of the halo is fairly
close to an isothermal profile. 

In the case of halos with masses comparable to that of rich
clusters, $M_{200} \sim 10^{15} \Msun$, the systematic error in 
the mass due to the assumption of an isothermal potential is
small.  Therefore, the masses of lensing clusters that are estimated
under the assumption of an isothermal potential (and in the limit
that the shear is detected out to a radius that is large enough
to be comparable to $r_{200}$) should not have large systematic
errors if, indeed, their density profiles are fitted well by NFW profiles.

However, recent observations of lensing of distant field galaxies
by nearby field galaxies (and, additionally, by the individual 
galaxies within clusters, e.g., Natarajan et al.\ 1998) have inspired
a number of investigations through which the mass and extent of the
dark matter halos of the lens galaxies might be constrained.  The
technique, known as galaxy--galaxy lensing, seems very promising at
the moment, and in the near future a considerable amount of effort
will be devoted to the use of observations of galaxy--galaxy 
lensing to constrain the nature of the dark matter halos of galaxies.
The results of our investigation of systematic errors in the
mass estimated for NFW lenses under the assumption 
of an isothermal
potential indicate that these errors can be significant for 
galaxy--mass lenses ($\sim 60\%$ 
in the case of a COBE-normalized CDM universe).  Therefore, in the
upcoming studies of galaxy--galaxy lensing, should an observational
constraint on the masses of galaxy halos be based upon the assumption
of an isothermal potential, 
it will be important
to keep such systematic errors in mind when judging the 
strength of such a constraint.

\section*{Acknowledgments}

Support under NSF contract
AST-9616968 (TGB and COW) and an NSF Graduate Fellowship
(COW) are gratefully acknowledged.

\clearpage

\begin{figure}
\plotfiddle{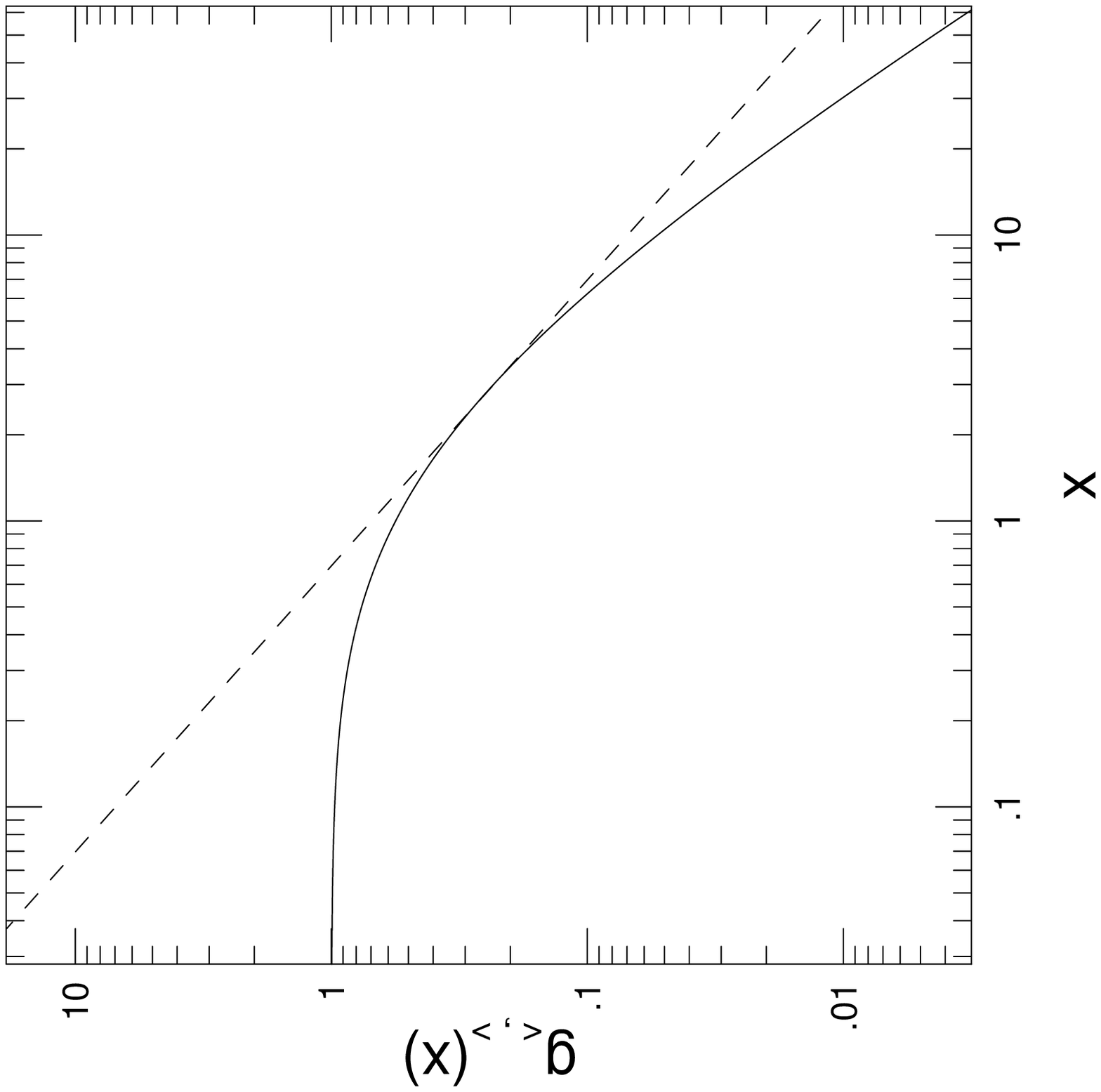}{5.0in}{0}{60.0}{60.0}{-250}{50}
\vspace{ -1.0in}
\caption{Solid line: 
the dependence of $\gamma_{\rm nfw}$ on the dimensionless
radius $x = R/r_s$, i.e., equations (\ref{gless}) and (\ref{ggtr}).  Also
shown for comparison (dashed line) is the radial dependence of
the shear due to a singular isothermal lens, $\gamma_{\rm sis}(x)
\propto x^{-1}$.  The normalization is arbitrary.}
\end{figure}

\clearpage
\begin{figure}
\plotfiddle{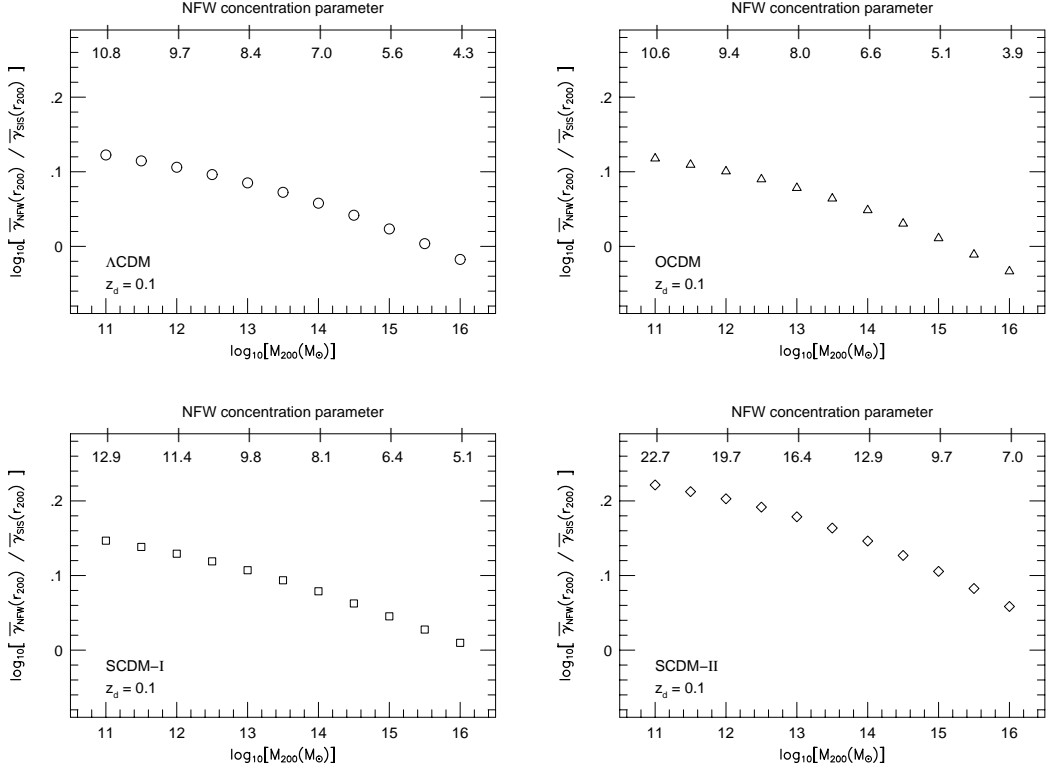}{5.0in}{0}{60.0}{60.0}{-250}{50}
\vspace{-1.0in}
\caption{Ratio of the mean shear interior to a radius of 
$r_{200}$ for NFW and SIS lenses.  Both lenses are constrained
to have identical masses interior to $r_{200}$ (see text).  The
panels correspond to four different CDM models, and in this figure
all of the lenses were placed at redshift of $z_d = 0.1$.}
\end{figure}

\clearpage
\begin{figure}
\plotfiddle{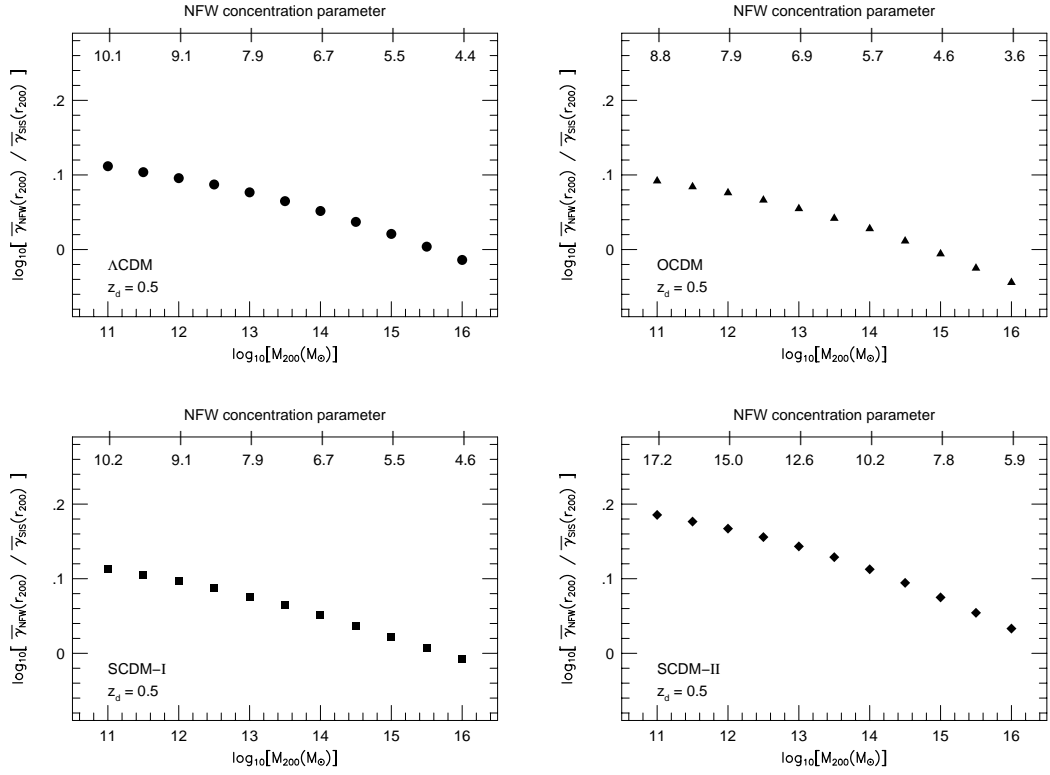}{5.0in}{0}{60.0}{60.0}{-250}{50}
\vspace{-0.6in}
\caption{Same as Fig.\ 2, but the lenses were placed at
a redshift of $z_d = 0.5$}
\end{figure}

\clearpage
\begin{figure}
\plotfiddle{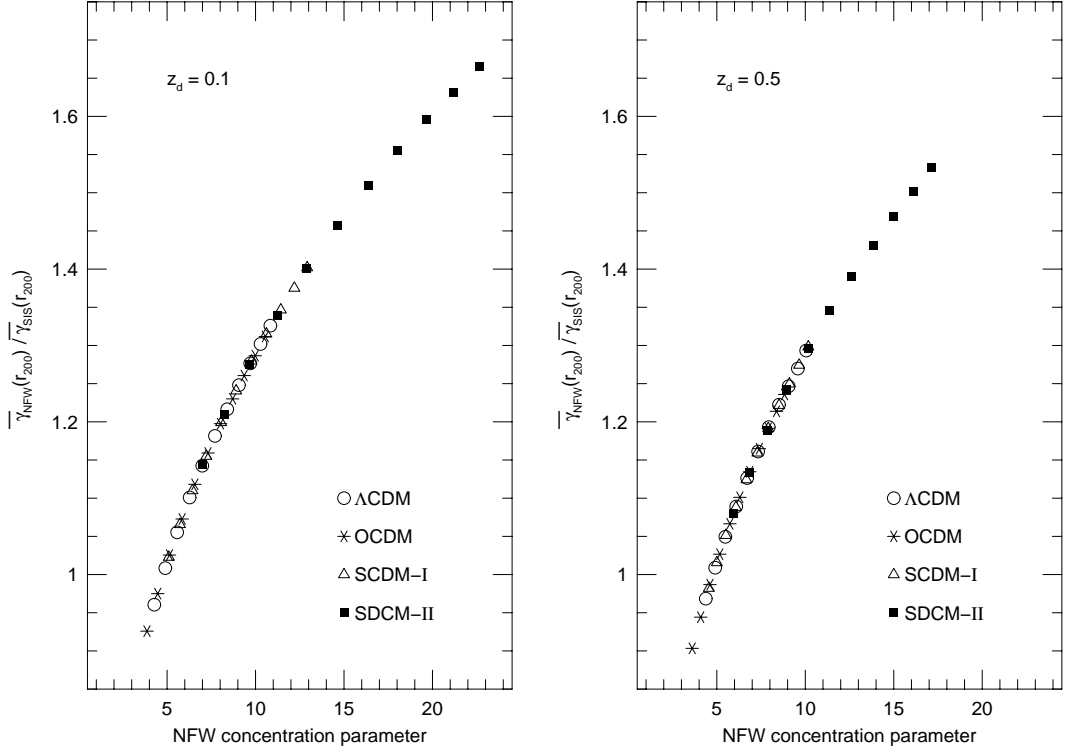}{5.0in}{0}{60.0}{60.0}{-250}{50}
\vspace{-0.6in}
\caption{Ratio of the mean shear interior to a radius of
$r_{200}$ for NFW and SIS lenses as a function of the halo
concentration parameter.  The point types refer to four different
CDM models.  The left panel shows the result obtained when all lenses
are placed at a redshift of $z_d = 0.1$; the right panel shows the result
obtained when all lenses are placed at a redshift of $z_d = 0.5$.}
\end{figure}

\end{document}